\documentclass[traditabstract]{aa}
\pdfoutput=1

\usepackage[varg]{txfonts}
\usepackage{graphicx}
\usepackage{natbib}
\usepackage{url}

\renewcommand{\hat}[0]{HAT-P-11~b}

\begin{document}

\title{Discovery of the secondary eclipse of \hat}
\author{K.~F.~Huber \and S.~Czesla \and J.~H.~M.~M.~Schmitt}
\institute{Hamburger Sternwarte, Universit\"at Hamburg, Gojenbergsweg 112, 21029
Hamburg, Germany}
\date{Received ... / Accepted ... }

\abstract{We report the detection of the secondary eclipse of \hat, 
a Neptune-sized planet orbiting an active K4~dwarf.
Using all available short-cadence data of the \textit{Kepler} mission,
we derive refined planetary ephemeris increasing their precision by more than an order of magnitude.
Our simultaneous primary and secondary transit modeling results in improved transit and orbital parameters.
In particular, the precise timing of the secondary eclipse allows to
pin down the orbital eccentricity to $0.26459_{-0.00048}^{+0.00069}$.
The secondary eclipse depth of $6.09_{-1.11}^{+1.12}$~ppm corresponds to a $5.5\sigma$ detection and
results in a geometric albedo of $0.39\pm0.07$ for \hat, close to Neptune's value,
which may indicate further resemblances between these two bodies.
Due to the substantial orbital eccentricity, the planetary equilibrium temperature is expected to 
change significantly with orbital position and ought to vary between $630^\circ$~K and $950^\circ$~K,
depending on the details of heat redistribution in the atmosphere of \hat.}

\maketitle

\section{Introduction}

The extrasolar planet \hat\ (also known as KOI-3.01 and  Kepler-3~b) 
is a super-Neptune orbiting an active K4~dwarf with a period of 4.89~days.
It was discovered through the transit method by \citet{Bakos2010}, using ground-based photometry 
in the context of the HAT project \citep{Bakos2002, Bakos2004}.
In combination with radial velocity (RV) measurements, \citet{Bakos2010} derived the orbital and
planetary parameters and found that \hat\ is on a rather eccentric orbit.
Using RV measurements obtained with HIRES \citep{Vogt1994},
\citet{Winn2010} confirmed the eccentric orbit and refined the orbital parameters of \hat.
Using the Rossiter-McLaughlin effect \citep{Rossiter1924,McLaughlin1924},
\citet{Winn2010} conclude that the planetary orbit is highly inclined and that 
the planet crosses the stellar disk almost parallel to the
sky-projected stellar rotation axis, a result
later confirmed by \citet{Hirano2011} with independent RV measurements.

\hat\ was also observed by the \textit{Kepler} mission \citep{Borucki2010}.
\citet{Deming2011} analyzed the
\textit{Kepler} light curve of quarters~$0$ to $2$ along with ground-based photometry
and present updated transit ephemeris and planetary parameters. 
The host-star HAT-P-11 is highly active, showing strong
rotational modulation attributable to starspots,
from which a stellar rotation period of about $29$~d has been inferred \citep{Bakos2010};
the potentially integer ratio of $6$:$1$ between the stellar rotation and the planetary orbital period
of \hat\ has nurtured quite a bit of discussion on
star-planet interactions in the system \citep{Beky2014}.
In addition to rotational modulation, prominent starspot-crossing signatures are
clearly visible in the majority of \textit{Kepler} transit light curves.
These features were discussed by several authors including \citet{Deming2011} and \citet{Beky2014}.
\citet{Sanchis-Ojeda2011} present an in-depth analysis of the spot-crossing features in 
$26$~\textit{Kepler} transits and conclude that the repetitive appearance of spot-occultations 
in similar locations in the transit profiles confirms the misalignment of the system and 
indicates active latitudes on the star.

A number of attempts to detect the secondary eclipse of \hat\ using the \textit{Kepler} 
photometry have been made \citep[e.g.,][]{Southworth2011, Angerhausen2015}, yet
a secondary eclipse  was not detected, and the same applies to the case of the planetary phase curve \citep{Angerhausen2015}.  
While \citet{Deming2011} mention the existence of \textit{Spitzer} observations of the secondary eclipse, neither a detailed analysis nor a detection has been published so far in the
refereed literature.

\textit{Kepler} stopped observing \hat\ in $2013$ after obtaining $15$~quarters or more than four years
of data.
In this paper we analyze the entire body of \textit{Kepler} photometry of \hat, comprising more than
two hundred primary transits, derive precise ephemeris and primary transit parameters, and
ultimately detect the secondary eclipse signature.

\section{Data analysis}

\subsection{Primary transit ephemeris}
\label{Sec:ana-prim}

The orbital parameters of \hat\ have been determined by several authors
based on different data sets \citep{Bakos2010,Winn2010,Sanchis-Ojeda2011}.
Since by now $15$~quarters of space-based \textit{Kepler} photometry are available,
of which only a fraction has previously been used to derive ephemeris and transit parameters,
we first determine updated values based on the entire available data set of \textit{Kepler}.

We specifically use the data from those $14$~quarters for which short-cadence photometry
is available, which are quarters 0, 1, 2, 3, 4, 5, 6, 9, 10, 12, 13, 14, 16, and 17.
These data cover a total of $222$~primary transits.
For quarter~$8$ only long-cadence data are available, which are not considered,
because the $14$~additional transits in that quarter do not significantly improve
our results, and further, deformations caused by spot-crossing events are much harder
to identify in long-cadence data, which makes
it more difficult to account for them in the analysis.
For the quarters~$7$, $11$, and $15$ no observations of \mbox{HAT-P-11}
are available in the MAST archive.

Our analysis is based on the simple aperture photometry (SAP) provided by the
\textit{Kepler} pipeline reduction from which we
removed all data points flagged as invalid or bad by the pipeline. 
Of the $222$~primary transits observed in
short-cadence mode, we exclude $16$ transits from our analysis, since they
lack proper pre- and post-transit coverage and can therefore not be reliably normalized;
we consider at least ten data points before and after the transit indispensable to
carry out an appropriate continuum normalization.

Prior to modeling the transits, we first divide the light curve of each quarter 
by its median flux.   We then identified the transits based on the 
orbital period~$P_\mathrm{ref}$ and reference epoch~$T_\mathrm{ref}$
given by \citet{Sanchis-Ojeda2011} and note that their reference epoch
is reported in BJD$_\mathrm{TDB}$, but \textit{Kepler} times are provided as
BJD$_\mathrm{UTC}$\footnote{\url{http://archive.stsci.edu/kepler/release_notes/release_notes19/DataRelease_19_20130204.pdf}, Sect.~$3.4$},
which lag behind the TDB system by $66.184$~seconds prior to quarter~$14$ and one
additional leap second, introduced during the first month of quarter~$14$.
TDB refers to ``Temps Dynamique Barycentrique'', a relativistic time system 
defined in the barycentric reference frame of the Solar System.
In our analysis, we rely on UTC times.

The ``continuum'' light curve surrounding each transit is fit with a first-order polynomial.
The light curve of \hat\ shows strong rotational modulation with an amplitude of up to $2$~\% caused 
by starspots.
To minimize the impact of the rotating spots on the transit normalization,
we adopt the normalization procedure described by \citet{Czesla2009}. Rather than
dividing by the continuum fit, we, first, subtract it and, second,
divide by the maximum brightness in the light curve ($f_\mathrm{max} = 1.01101545$),
estimated from the highest peak in our
median-normalized light curve. In this last step, we implicitly assume that the stellar
brightness of \hat\ shows no long-term trend during the \textit{Kepler} observations.

As discussed by \citet{Sanchis-Ojeda2011} and \citet{Beky2014},
a significant number of transits is strongly affected by spot-crossing features.
To evaluate the degree to which individual transits are deformed by such features,
we use a spot-free transit model based on the approach by \citet{Mandel2002},
perform a Nelder-Mead simplex fit to each transit \citep{Nelder1965}, and determine the in-transit $\chi^2_t$ value;
the error of data points is estimated by their standard deviation around the fit in the continuum.
As we are currently only concerned with transit timing, we do not consider the orbital 
eccentricity in the transit modeling; our estimates show that
the difference between ingress and egress duration is only~$0.1$~seconds,
making the deviation from a symmetrical transit shape negligible.
Because we fix the orbital period to the value reported by \citet{Sanchis-Ojeda2011} in this step,
our fit adjusts the transit duration adopting unphysical
values for the semi-major axis, which we ignore in our analysis.

\begin{figure*}[t]
  \includegraphics[width=0.95\textwidth]{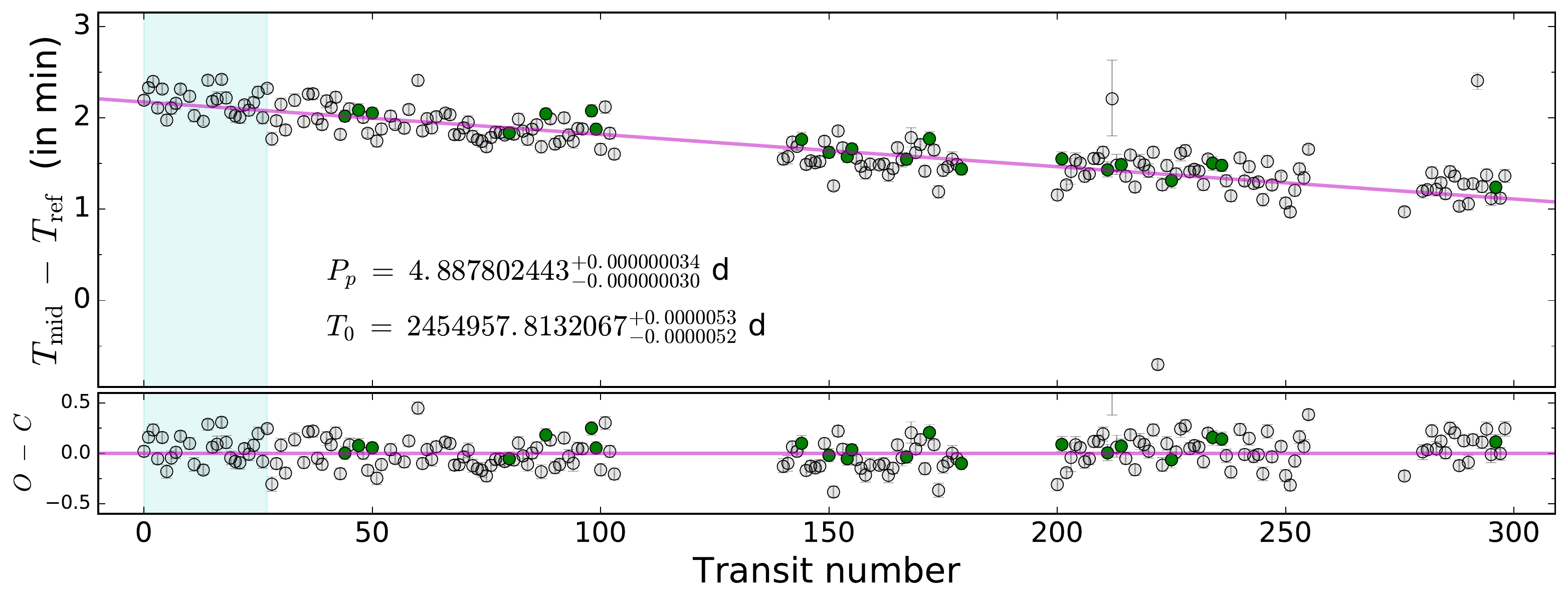}
  \caption{Measurements of mid-transit times from $206$~\textit{Kepler} transits.
           The green filled circles are our selection of $21$~low-$\chi^2$ transits.
           The line represents a first-order polynomial fit to the error-weighted measurements.
           The lower panel shows the residuals with outliers cut-off;
           the variations are primarily caused by deformations of the transit due to starspots.
           The shaded area contains the transits that previously have been analyzed by \citet{Sanchis-Ojeda2011}.
           See Sects.~\ref{Sec:ana-prim} and~\ref{Sec:res-eph} for detailed explanations.
  \label{fig:T0s}}
\end{figure*}

Based on these fits, we identify the lowest decile of the transits ($21$) with the smallest $\chi^2_t$ values ($0.9 \le \chi^2_\mathrm{red} \le 1.3$), which we consider to show the weakest spot contamination. 
Based on this subsample, we re-determine the transit parameters and their errors with a MCMC sampling approach\footnote{\url{https://github.com/pymc-devs/pymc}}.
In particular, the $21$~transits are simultaneously fit with a model where orbital inclination~$i$, planet-to-star radius ratio~$R_p/R_s$,
scaled semi-major axis~$a/R_s$, and linear and quadratic limb-darkening coefficients~$u_1$ and $u_2$ are coupled,
and only the mid-transit times are varied individually. Again,
the orbital period~$P_p$ remained fixed at the value reported by \citet{Sanchis-Ojeda2011}.
The $21$~individual results for the mid-transit times~$T_\mathrm{mid}$ are plotted in Fig.~\ref{fig:T0s} (filled green circles).
Unless explicitly stated otherwise, we give the median of the marginal posterior 
distributions determined using the MCMC sampling
as the parameter estimate.  Our
uncertainties refer to the $68$~\% credibility intervals defined by the $16$~\% and $84$~\% quantiles.

Mid-transit times for
the remaining $185$~transits were determined by modeling them with all transit
parameters but the mid-transit time kept fixed at the previously determined values.
The resulting mid-transit times are also shown in Fig.~\ref{fig:T0s} 
(open circles).~From all $206$~measurements of $T_\mathrm{mid}$ we derive revised ephemeris $T_0$ and $P_p$
presented in Table~\ref{Tab:T0-transit}.
More details and a discussion of these results are provided in Sect.~\ref{Sec:res-eph}.

\newcommand\Tstrut{\rule{0pt}{2.3ex}}
\renewcommand{\arraystretch}{1.5}
\begin{table}
\caption{Reference and revised ephemeris}
\label{Tab:T0-transit}
\centering
\begin{tabular}{lr}
 \hline \hline
 Parameter                                                 & Value\Tstrut \\ \hline 
 Reference ephemeris$^a$                                   & \\ \hline
 Reference epoch $T_\mathrm{ref}$ (BJD$_\mathrm{UTC}$)$^b$ & $2454957.8116980$ \\
 Reference period $P_\mathrm{ref}$ (days)                  & $4.8878049$ \\ \hline\hline
 Revised ephemeris$^c$                                     & \\ \hline
 Mid-transit time $T_0$ (BJD$_\mathrm{UTC}$) & $2454957.8132067^{+0.0000053}_{-0.0000052}$\Tstrut \\
 Orbital period $P_p$ (days)                 & $4.887802443^{+0.000000034}_{-0.000000030}$ \\ \hline
\end{tabular}
\tablefoot{$^a$ Taken from \citet{Sanchis-Ojeda2011}.
$^b$ Converted into UTC by subtracting $66.184$~s.
$^c$ Computed from the measurements presented in Fig.~\ref{fig:T0s}.
}
\end{table}

\subsection{Secondary eclipse}
\label{Sec:ana-sec}

Next we use our updated ephemeris to phase-fold the light curve and search for the secondary transit of \hat.
We define the phase~$p$ as
\begin{equation}
p = \frac{t \, - \, T_0}{P_p} \, - \, \left\lfloor \frac{t \, - \, T_0}{P_p} \right\rfloor \ ,
\label{Eq:phase}
\end{equation}
where the brackets around the second term represent the floor function.
Since the orbit is not circular, the secondary transit is not necessarily at phase~$0.5$.
Using the known planetary orbit elements \citep{Winn2010}, we estimate the duration of the
secondary transit to be at least $0.024$ in phase.
While in principle also the timing data can be derived from the orbit elements,
we first carry out a search for the secondary transit pretending ignorance of the timing 
information.

In particular, we assume a series of $200$ hypothetical mid-transit phases between
$p=0.25$ and $p=0.75$ with a phase spacing of $0.0025$.
For each of the $200$~mid-transit phases, we
normalize all observed (hypothetical) transits by dividing by a first-order polynomial fit
to the adjacent continuum and superimpose the phase-folded, normalized light curves.
To avoid including the secondary transit in the continuum normalization,
we define a continuum range of $[p-0.025, p-0.013]$ and $[p+0.013, p+0.025]$ around mid-transit
phase, which leaves some margin for a secondary transit duration of up to $0.026$.
Again, transits with insufficient continuum coverage were rejected.
The superposed, normalized light curves were rebinned to a binsize of $0.001$ in phase, which
yields approximately $1500$ data points per bin.
To estimate the error of the rebinned data points, we calculate the standard deviation of points in all $200$ hypothetical secondary transits,
resulting in a value of \mbox{$\sigma_\mathrm{reb} = 3.72 \cdot 10^{-6}$}.

\begin{figure}[t]
  \includegraphics[width=0.49\textwidth]{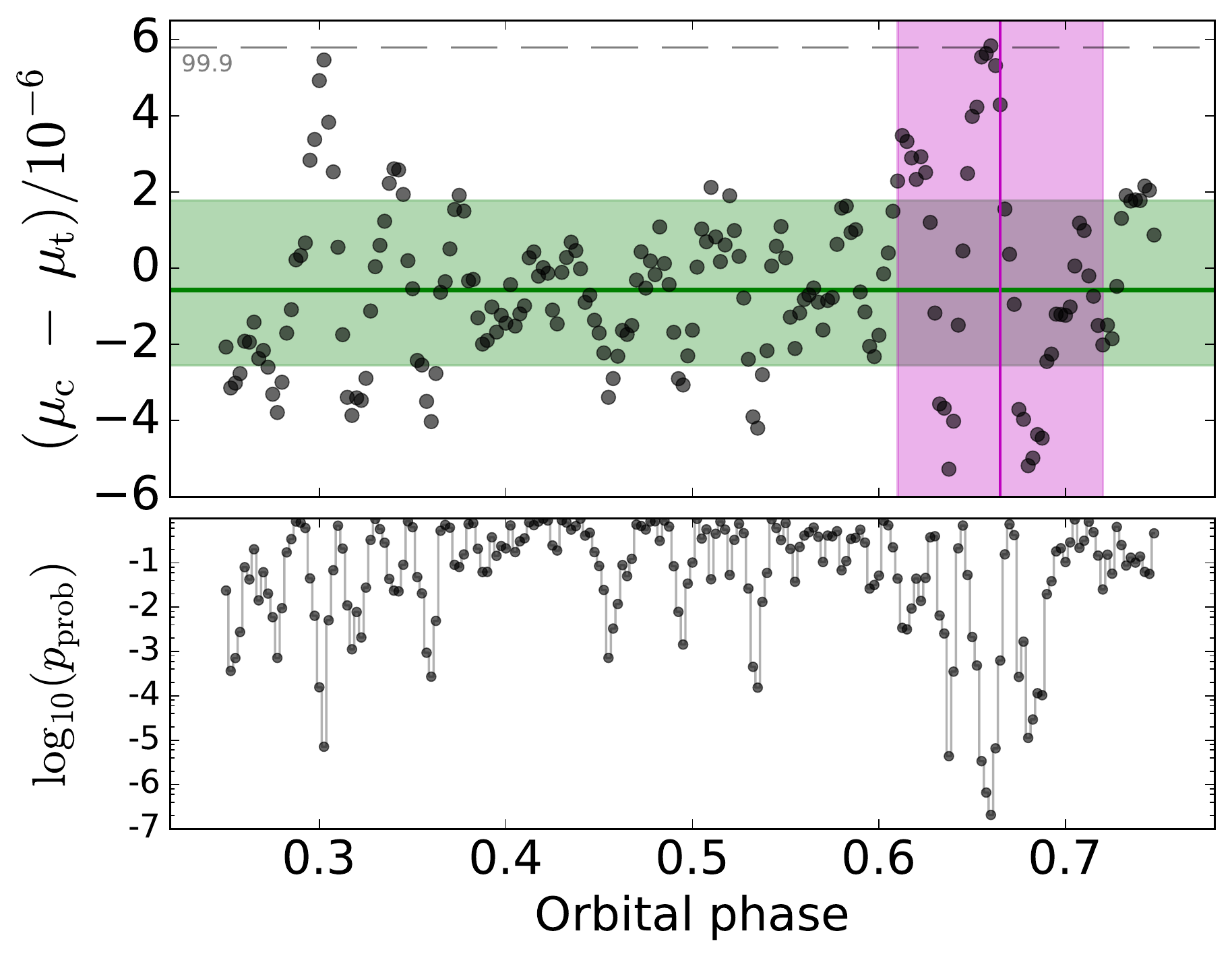}
  \caption{\textit{Upper panel}: Mean of continuum points~$\mu_c$ minus mean of in-transit points~$\mu_t$
           over orbital phase (see Sect.~\ref{Sec:ana-sec} for details).
           The horizontal (green) area marks the range between the $16$~\%
           and $84$~\% quantiles of the distribution of points,
           the gray dashed line indicates the $99.9$~\% quantile.
           The position of the secondary transit of \hat\ predicted by RV measurements
           is given by the vertical (magenta) line;
           the shaded area indicates the $\pm \sigma$ uncertainty of the prediction.
           \textit{Lower panel}: Estimate of the significance of the measurements in the upper panel
           using an F-test (see Sect.~\ref{Sec:ana-sec} for details).
  \label{fig:mean-diff}}
\end{figure}

For each hypothetical transit we then calculate the mean of the continuum points $\mu_c$
and the mean of in-transit points $\mu_t$.  For those choices of mid-transit phase that
do not correspond to a real physical signal,
the derived secondary transit depth \mbox{$\mu_c - \mu_t$}
will assume stochastic values determined by
the characteristics of the light curve. However, any detectable transit-like signal
should be associated with a significant positive excursion.
In Fig.~\ref{fig:mean-diff}  (upper panel) we plot the derived $\mu_c - \mu_t$ values as a function of the
assumed mid-transit orbital phase.
We find two mid-transit phases for which the depth
strongly deviates toward the positive side of the distribution. 
The highest positive deviation is found at about phase~$0.66$.
Another, somewhat less pronounced peak is located at a phase of about $0.3$.
Both positive peaks are accompanied by negative deviations on either side;
these ``swings'' reflect the fact that points in the phase window move from the continuum region 
into the transit window and back into the continuum again as the hypothetical transit mid-point advances.
In fact, such a swing is expected from a persistent, transit-like feature in the light
curve with its true position located close to the peak.

To estimate the significance of the hypothetical transit signals presented in Fig.~\ref{fig:mean-diff} (upper panel), we apply the ANOVA F-test \citep{Rawlings2001}. Specifically, we 
compare the null hypothesis of a single mean flux value in the selected continuum and transit window with the
hypothesis of deviating fluxes $\mu_t$ and $\mu_c$ in the transit and continuum windows (i.e., a non-vanishing
transit depth) for each of our $200$~mid-transit phases. To that end, we compute the best-fit reduced $\chi^2$ values
for both models, compute the F-statistic, and obtain the $p$-value, which we show in the lower panel of
Fig.~\ref{fig:mean-diff}. The $p$-value gives the probability of obtaining an F-value as large or larger by chance
given that the null hypothesis is true, which means that the flux is indeed constant.
The largest hypothetical secondary transit depth at a phase of $0.66$ is associated with a $p$-value of nearly
$10^{-7}$, which is more than ten times smaller than for any other tested transit phase. Nonetheless, also
other structures are associated with low $p$-values, most notably, 
the peak at phase $\approx 0.3$.
We cannot rule out that this indicates another transit-like signal 
potentially originating in the HAT-P-11 system;
if so, we are unable to provide a reasonable explanation for its origin.
At any rate, the signal at phase $p \approx 0.66$ remains the deepest and most significant and,
as we will see shortly, is almost certainly associated with the planetary eclipse.

Based on the orbital elements reported by \citet{Winn2010}, we can derive the timing
of the secondary transit as predicted by the radial velocity analysis; note that we
shifted the argument of periastron, $\omega$, by
$180^\circ$ to represent the planetary rather than the stellar orbit \citep[e.g.,][]{Deming2011}.
The vertical line in Fig.~\ref{fig:mean-diff} (upper panel) gives the expected phasing of the secondary transit and the
magenta-colored, vertical area shows the range of uncertainty 
considering the errors in the eccentricity and the argument of periastron reported by \citet{Winn2010}.

Figure~\ref{fig:mean-diff} clearly demonstrates that the location of the deepest transit-like signal 
detected in our search is entirely consistent with the secondary transit phasing  
expected from the orbital elements reported by \citet{Winn2010}.
Therefore, we identify this feature with the secondary transit signature of \hat.
Our analysis yields an approximate mid-transit phase of $p=0.659$ for the secondary transit.
We use this together with the previously defined continuum range to normalize the data and
display the resulting rebinned light curve of the secondary eclipse 
in Fig.~\ref{fig:transits} (right panel).

\subsection{Eccentric orbit}

To carry out a more detailed analysis of the transits,
we now model both the primary and secondary transit simultaneously.
The primary transit modeling is based on the $21$~transits with weak starspot
contamination identified in Sect.~\ref{Sec:ana-prim}, which we phase-folded with our orbital period.
Each transit is again normalized individually according to the procedure described in Sect.~\ref{Sec:ana-prim}.
For the secondary transit, the preliminary mid-transit time, binsize, and continuum are defined according to Sect.~\ref{Sec:ana-sec},
and we adopt $\sigma_\mathrm{reb}$ as error for each point of the rebinned light curve. 
All transit parameters, now including the eccentricity~$e$,
the argument of periastron~$\omega$,
and the time of periastron passage~$\tau$ are varied
and their posterior distributions are sampled by the MCMC algorithm.

We model the secondary transit using a ``conventional'' primary transit model with no
stellar limb darkening and introduce an additional parameter for the brightness
contrast between the stellar and planetary disk.
With the brightness contrast described by the parameter~$f_3$
and a first-order polynomial parameterization of the continuum
with offset~$c_o$ and gradient~$c_l$,
the observed flux~$f$ of the secondary transit light curve is modeled by the function
\begin{equation}
f(t) \, = \, \left( 1 \, - \, \frac{df(t)}{1+f_3} \right) \, \left( c_o \, + \, c_l\, t \right)^{-1} \ ,
\end{equation}
where $df(t)$ is the relative decrease in flux calculated by the geometric transit model
\citep{Mandel2002}.
All other parameters are coupled to the values of the primary transit so that both transits
are modeled based on the same set of orbital elements.
For technical reasons, we applied a shift of $180^{\circ}$ to the argument of periastron, which
effectively reflects the orbit through the origin, allowing to model the secondary transit as a primary transit.

\begin{figure*}[t]
  \includegraphics[width=0.99\textwidth]{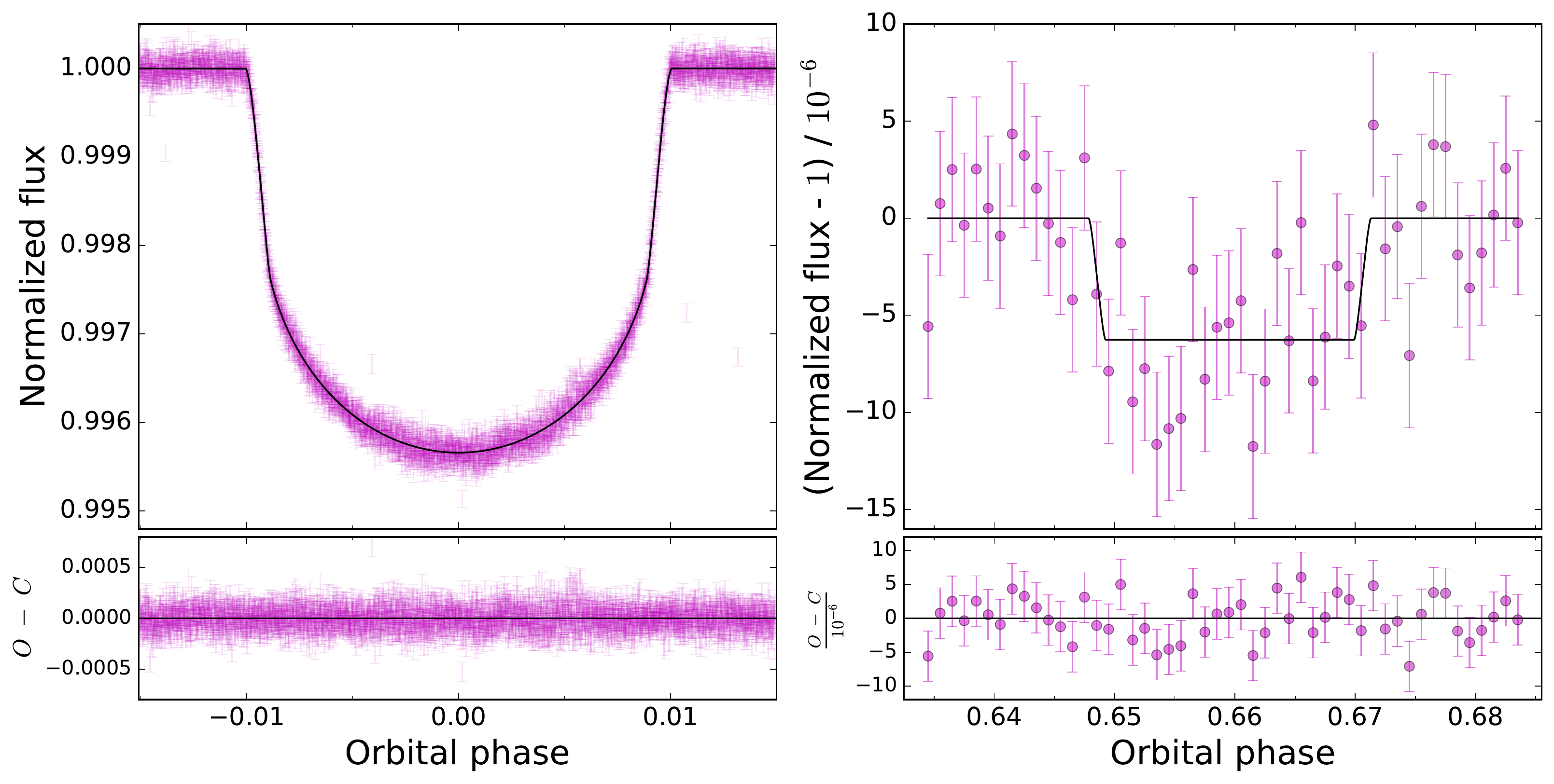}
  \caption{Data of the primary (left panel) and secondary transit (right panel) overplotted by the lowest deviance solution (black line).
           The lower panels show the residuals. The primary data contains almost $4500$~data points coming from the sample of $21$~low-$\chi^2$ transits;
           the secondary data is rebinned to intervals of length $0.001$ in phase, each containing roughly $1500$~short-cadence points.
  \label{fig:transits}}
\end{figure*}

In a first step, we use uniform priors in the modeling to derive estimates based solely on the \textit{Kepler} photometry.
In order to incorporate prior knowledge, we then repeat the MCMC sampling
including Gaussian priors on $R_p/R_s$, $i$,
$a/R_s$, $e$, and $\omega$ based on the values reported by \citet{Winn2010}.
The prior information comprises results derived from RV measurements,
which usually constrain the eccentricity and argument of periastron better than photometry alone.
The results are given in Table~\ref{Tab:transitsFit}, also containing the lowest deviance solution
which is plotted in Fig.~\ref{fig:transits} along with the data.

\renewcommand{\arraystretch}{1.5}
\begin{table*}
\caption{Model parameters of \hat.}
\label{Tab:transitsFit}
\centering
\begin{tabular}{lrrr}
 \hline \hline
 Parameter  & Free (uniform priors) & Gaussian priors$^a$ & Lowest deviance solution \\ \hline 
                       Planet-to-star radius ratio $R_p/R_s$ & $           0.05856_{-0.00015}^{+0.00013}$ & $           0.05850_{-0.00013}^{+0.00009}$ & $                                 0.05854$ \\
                              Scaled semi-major axis $a/R_s$ & $                   14.64_{-0.09}^{+0.10}$ & $                   14.68_{-0.06}^{+0.09}$ & $                                   14.64$ \\
                               Orbital inclination $i$ (deg) & $                   88.99_{-0.13}^{+0.17}$ & $                   89.05_{-0.09}^{+0.15}$ & $                                   89.00$ \\
                     Linear limb-darkening coefficient $u_1$ & $                 0.646_{-0.008}^{+0.008}$ & $                 0.645_{-0.007}^{+0.008}$ & $                                   0.646$ \\
                  Quadratic limb-darkening coefficient $u_2$ & $                 0.048_{-0.015}^{+0.016}$ & $                 0.051_{-0.015}^{+0.014}$ & $                                   0.049$ \\
                                Primary impact parameter $b$ & $                 0.222_{-0.035}^{+0.026}$ & $                 0.209_{-0.032}^{+0.019}$ & $                                   0.220$ \\
                   Primary transit duration $T_{14}$ (hours) & $              2.3572_{-0.0017}^{+0.0017}$ & $              2.3565_{-0.0016}^{+0.0015}$ & $                                  2.3573$ \\ \hline
                                    Orbital eccentricity $e$ & $           0.26493_{-0.00091}^{+0.00033}$ & $           0.26459_{-0.00048}^{+0.00069}$ & $                                 0.26528$ \\
                       Argument of periastron $\omega$ (deg) & $              -162.149_{-0.086}^{+0.043}$ & $              -162.226_{-0.094}^{+0.203}$ & $                                -162.157$ \\
                   Time of periastron passage $\tau$ (phase) & $           0.87439_{-0.00023}^{+0.00016}$ & $           0.87421_{-0.00010}^{+0.00036}$ & $                                 0.87451$ \\ \hline
          Separation of transits $\Delta t_\mathrm{s-p}$ (phase) & $           0.65933_{-0.00054}^{+0.00023}$ & $           0.65909_{-0.00032}^{+0.00059}$ & $                                 0.65955$ \\
                   Secondary transit depth $d_s$ ($10^{-6}$) & $                    6.14_{-1.10}^{+1.10}$ & $                    6.09_{-1.11}^{+1.12}$ & $                                    6.26$ \\
                         Brightness contrast parameter $f_3$ & $                        557_{-84}^{+122}$ & $                        560_{-87}^{+124}$ & $                                     546$ \\
                 Secondary impact parameter $b_\mathrm{sec}$ & $                 0.261_{-0.041}^{+0.031}$ & $                 0.246_{-0.038}^{+0.022}$ & $                                   0.259$ \\
    Secondary transit duration $T_{14,\mathrm{sec}}$ (hours) & $              2.7484_{-0.0056}^{+0.0064}$ & $              2.7503_{-0.0046}^{+0.0053}$ & $                                  2.7502$ \\ \hline
                            Linear normalization coefficient & $       -0.000024_{-0.000023}^{+0.000018}$ & $       -0.000044_{-0.000033}^{+0.000043}$ & $                               -0.000040$ \\
                            Offset normalization coefficient & $        1.000013_{-0.000012}^{+0.000015}$ & $        1.000026_{-0.000028}^{+0.000022}$ & $                                1.000023$ \\
\hline
\end{tabular}
\tablefoot{$^a$ Values and uncertainties of $R_p/R_s$, $a/R_s$, $i$, $e$, and $\omega$ are taken from \citet{Winn2010} and used as input for Gaussian priors.
All other parameters have uniform priors.}
\end{table*}

\section{Results}

\subsection{Ephemeris}
\label{Sec:res-eph}

From our modeling presented in Sect.~\ref{Sec:ana-prim} we can derive estimates for the mid-transit times
\begin{equation}
T_c = T_1 \, + \, \frac{T_4 - T_1}{2} \ ,
\end{equation}
for all $206$~considered primary transits in the \textit{Kepler} light curve; here,
$T_1$ and $T_4$ denote the start of ingress and the end of egress.
We compare the thus derived mid-transit times to the ephemeris
$T_\mathrm{ref}$ and $P_\mathrm{ref}$ (Table~\ref{Tab:T0-transit})
given by \citet{Sanchis-Ojeda2011}
by calculating the offset according to
\begin{equation}
T_{c, E}  - (T_\mathrm{ref} + E\, P_\mathrm{ref}) = T_\mathrm{mid} - T_\mathrm{ref} \ ,
\end{equation}
where $E$ is the epoch with respect to the first \textit{Kepler} transit \mbox{($E=0$)}.
The resulting offsets are given in Fig.~\ref{fig:T0s}.

At early epochs, there is a clear offset of about $2$~minutes between our measurements
and the ephemeris given by \citet{Sanchis-Ojeda2011}, which we speculate results
from a different definition of their reference time, $T_\mathrm{ref}$, and our mid-transit time.
As the epoch advances, the offset decreases, indicating that the orbital period is slightly shorter
than the reference period reported by \citet{Sanchis-Ojeda2011}.

We calculate updated ephemeris
by fitting the measurements with a first-order polynomial (linear function) using MCMC sampling.
The straight line in Fig.~\ref{fig:T0s} indicates
the lowest deviance solution and the parameter estimates are given in Table~\ref{Tab:T0-transit}.
Repeating the fit with a Nelder-Mead minimization algorithm \citep{Nelder1965} produces results
numerically identical to the MCMC lowest deviance solutions.
We define $T_0$ as the expected mid-transit time at epoch zero,
which equals the median value of the offset in our polynomial fit.
While the derived orbital period is somewhat lower than that reported by
\citet{Sanchis-Ojeda2011}, who studied only the first $28$~transits
(shaded area of Fig.~\ref{fig:T0s}), it is still consistent within their $2\sigma$ uncertainty.
Extrapolating our ephemeris backward to the epoch corresponding to the central
transit time given by \citet{Bakos2010},
we obtain an offset of $1.06$~s, which is entirely consistent with
their stated uncertainty of $28$~s.

Figure~\ref{fig:T0s} also shows the residuals ($O-C$) in the lower panel.
Clearly, the statistical errors of the individual transit mid-times are smaller than the
width of their distribution.
Since the shift in individual mid-transit times also depends
on the deformation of the transit profile by spot-crossing events \citep{Ioannidis2016},
we attribute this discrepancy to the effects of stellar activity.

Although our measurements do pin down the statistical uncertainty on $T_0$ to less than one second and
to less than $10^{-2}$~s for the orbital period,
one should keep in mind that these statistical estimates might be
somewhat optimistic in view of the pronounced transit deformations caused by starspot occultations.
With $206$~measurements, however, we expect that the orbital period should only weakly be affected
by the spot crossings, even if their positions are not statistically distributed in the transits
as found by \citet{Sanchis-Ojeda2011} and \citet{Beky2014}.
However, with spot features repeating at similar locations, which might be the case for \hat\
during all of the \textit{Kepler} observations, $T_0$ could be systematically shifted.
To obtain some estimate of the possible magnitude of that effect,
we check how much our reference time shifts (keeping $P_p$ fixed)
if we only fit the $21$~low-$\chi^2$ transits (filled, green circles in Fig.~\ref{fig:T0s}).
This results in an offset of about $3.5$~seconds, which is roughly ten times the statistical uncertainty
of $T_0$ and probably represents a more realistic estimate of the true uncertainty.

\subsection{Primary transit modeling}

\subsubsection{Planetary radius}

One of the key results of photometric transit modeling is the planet-to-star radius ratio.
\citet{Deming2011} report stellar and planetary radii of
\mbox{$R_s = (0.683 \pm 0.009) \, R_\odot$} and \mbox{$R_p = (4.39 \pm 0.06) \, R_\oplus$}.
In their analysis,
they estimate that uneclipsed starspots, not prominently showing as crossing events during the transit,
make the planetary radius appear larger by $1.76$~\%,
reducing the derived physical planetary radius to a value of $(4.31 \pm 0.06) \, R_\oplus$.

Adopting their stellar radius, and relying on our measurement of $R_p/R_s$ (Table~\ref{Tab:transitsFit})
from our sample of $21$ low-$\chi^2$ transits,
we determine a planetary radius of $(4.36\pm0.06) \, R_\oplus$.
As a result of our normalization procedure (see Sect.~\ref{Sec:ana-prim}),
we expect the effects of unocculted spots on our planet-to-star radius ratio
to be small, and we do not apply \citeauthor{Deming2011}'s correction factor.
Our result for the planetary radius is slightly larger,
but still consistent with the spot-corrected value reported by \citet{Deming2011}.

A systematic error in the planetary radius derived in our analysis may result 
from persistent and symmetrical spot coverage not showing in the rotational modulation (e.g., a long-lived
polar spot), which
can neither be excluded by our analysis nor would it be accounted for by our normalization approach.
While such a configuration would lead to an overestimation of the radius,
we expect that our sample of low-$\chi^2$ transits is also influenced by spot-crossing features at some level.
In fact, the distribution of residuals with respect to the primary transit model (Fig.~\ref{fig:transits}, lower left panel) is symmetrical.
However, a reduced $\chi^2$ value of $2.4$ resulting from the primary transit model,
leads us to the assumption that
the error of individual in-transit data points is underestimated and is larger than in the
adjacent continuum, from where we obtain the error estimate.
Additionally, a likely signature of a spot-crossing feature is visible at orbital phase $\approx 0.005$.
This indicates that additional noise sources affect the in-transit light curve, most likely including
signatures of small, unresolved spot-crossing events.
Such signatures lead to an underestimation of the radius ratio, and thus
counteracts the effects of persistent, unocculted spots.
Therefore, 
we adopt $R_p =  (4.36\pm0.06) \, R_\oplus$ as our best estimate,
but caution that the uncertainty is likely underestimated because of the presence of
unaccounted for systematic errors.

\subsubsection{Limb darkening}

In general, stellar limb-darkening (LD) is considered a nuisance parameter in transit modeling
which has to be accounted for when deriving planetary parameters.
However, space-based (short-cadence) photometry of exoplanetary transits, as provided for example by \textit{Kepler} or \textit{CoRoT},
offers one of only very few possibilities to directly measure the LD of stellar disks with high precision.
There is an ongoing discussion on which limb-darkening ``laws'' should be used and whether one
achieves more reliable results when keeping the LD coefficients fixed to theoretically predicted values or not
\citep[e.g.,][]{Howarth2011, Csi2013, Mueller2013, Espinoza2015}.

We applied a quadratic limb-darkening law and left the LD coefficients free in the MCMC sampling
resulting in \mbox{$u_1 = 0.645_{-0.007}^{+0.008}$} and \mbox{$u_2 = 0.051_{-0.015}^{+0.014}$} for the linear and quadratic coefficient, respectively.
\citet{Claret2012} determine theoretical values from PHOENIX models for the \textit{Kepler} bandpass
using two different fitting approaches which yield parameter estimates between $0.6937$ and $0.7076$ for $u_1$
and $0.0382$ and $0.0772$ for $u_2$ ($T_\mathrm{eff} = 4800$~K and $\log g=4.5$).
For the latter, our result lies well between the two predictions;
considering its absolute value, the linear coefficient does not strongly deviate from the predicted range,
although within its uncertainties it lies significantly below the theoretical values.
The LD of an highly active star as HAT-P-11 is likely influenced by cool and hot regions on its surface
which is not considered in the theoretical predictions.
Also our determined values for $u_1$ and $u_2$ are probably affected by starspots to some degree.
Considering these systematic uncertainties, we conclude that our LD values agree reasonably with the theoretical expectation.

\subsection{Secondary eclipse and orbital eccentricity}
\label{Sec:res-sec}

Figure~\ref{fig:transits} presents the data of primary and secondary transit as well as the
lowest deviance model of the combined fit.
Table~\ref{Tab:transitsFit} contains the lowest deviance values,
the median values of the posterior distributions for a model with uniform priors,
and another model with the results of \citet{Winn2010} incorporated as Gaussian priors.
The parameter estimates from both models agree within their $68$~\% credibility intervals,
which also cover the lowest deviance solution.
This indicates that there is no strong influence of the prior information on the posterior distribution.
Actually, we find the width of the Gaussian priors to be much broader than the posterior credibility intervals
(e.g., almost $100$~times larger for $e$ and about $50$~times larger for $\omega$)
and conclude that the likelihood entirely dominates the posterior and,
thus, both solutions become virtually identical.
Nonetheless, as the estimated values including the prior information comprehensively represent the
available knowledge, we prefer this solution and base our following calculations and
discussions on these values.


For the secondary transit model, we obtain a reduced $\chi^2$ value of $0.9$
(Fig.~\ref{fig:transits}, lower right panel), indicating that errors may be slightly overestimated here.
In fact, the evolution of the in-transit flux of the secondary eclipse seems to show a positive gradient with
the minimum flux reached shortly after ingress. If this evolution should be real, we argue that it
can hardly be related to the secondary planetary eclipse, because
during its totality phase the planet is behind the star and not visible.
Recalling that the shown data set is composed of
$208$ individually normalized light curves, we argue that this structure is likely an artifact.

We calculate the Poisson noise $\sigma_\mathrm{limit}$ of each bin of the secondary transit data
for comparison to our estimated uncertainty of \mbox{$\sigma_\mathrm{reb} = 3.72 \cdot 10^{-6}$}.
Using an average count rate of \mbox{$2.7 \cdot 10^6$~e$^-$/s} (electrons per second) for the \textit{Kepler} light curve,
the length of each rebinned time interval ($421.6$~s),
and the number of averaged secondary transits ($208$),
we derive $\sigma_\mathrm{limit} = 2 \cdot 10^{-6}$.
Our estimated noise is only about a factor two larger than the theoretical limit.

According to our modeling, the depth $d_s$ of the secondary transit is $6.09_{-1.11}^{+1.12}$~ppm, which
corresponds to a $5.5\sigma$ detection of a decline in flux.
Therefore, we argue that it is extremely unlikely to be a statistical artifact, especially because
its position was correctly predicted by independent measurements of the orbital elements.
Nonetheless, the transit signal remains weak compared to the uncertainty of
individual data points.
\citet{Angerhausen2015} searched for the secondary eclipse concluding that the depth of the secondary
transit has to be less than $147$~ppm, which is of course consistent
with our results.

\subsubsection{Secondary transit duration}

The duration of the secondary transit is directly related to the orbital elements.
Table~\ref{Tab:transitsFit} provides calculations of the impact parameters~$b$ and $b_\mathrm{sec}$
of the primary and secondary transit as well as their durations $T_{14}$ and $T_{14,\mathrm{sec}}$.
The values of impact parameter and duration are determined using
\begin{equation}
b \, = \, a \cos(i) \left( \frac{1-e^2}{1 \pm e \sin\omega} \right)
\label{Eq:impact-par}
\end{equation}
and
\begin{equation}
t_\mathrm{dur} \, = \, \frac{P_p}{\pi} \arcsin \left( \frac{R_s}{a} \frac{\sqrt{\left( 1 + \frac{R_p}{R_s} \right)^2 - b^2}}{\sin i} \right) \frac{\sqrt{1 - e^2}}{1 \pm e \sin \omega} \ .
\label{Eq:trans-dur}
\end{equation}
Due to our definition of $\omega$ relating to the planet,
Eqs.~\ref{Eq:impact-par} and \ref{Eq:trans-dur} refer to the secondary eclipse for the plus sign
and to the primary eclipse for the minus sign wherever $\pm$ is given;
naturally, shifting $\omega$ by $180^\circ$ produces the same sign change.

%
%

To calculate the time between primary and secondary eclipse, we
use the equation
\begin{equation}
\frac{\Delta t_\mathrm{2-1}}{P_p} \, = \, \frac{1}{2 \pi \sqrt{1-e^2}}
\int_{\nu_1(t_1)}^{\nu_2(t_2)}
\left( \frac{1-e^2}{1+e\cos \nu} \right)^2 \, d\nu \ ,
\label{Eq:tsp}
\end{equation}
which yields the time between two points $t_1$ and $t_2$ of a planet on an eccentric orbit;
here $\nu$ denotes the true anomaly which is a function of time.
For the primary and secondary transits we use $\nu_1 = -\pi/2-\omega = \nu_\mathrm{p}$ and $\nu_2 = \pi/2-\omega = \nu_\mathrm{s}$
and numerically solve the equation for our MCMC chains of the eccentricity and the argument of
periastron, thus obtaining an uncertainty estimate for $\Delta t_\mathrm{s-p}/P_p$ with
a negligible numerical error. 
We note that the commonly used expression to derive the
separation in orbital phase between primary and secondary transit 
\begin{equation}
\frac{\Delta t_\mathrm{s-p}}{P_p} \, \approx
\, 1\, - \, \frac{1}{2 \pi} \left( \pi + 4 e \cos \omega \right) \ ,
\label{Eq:tsp_approx}
\end{equation}
which is the first-order Taylor series expansion of Eq.~\ref{Eq:tsp} at $e=0$,
should not be used in this case.\footnote{The subtraction from unity on the right-hand side of Eq.~\ref{Eq:tsp_approx}
is again a result of our shift of $\omega$ by $\pi$.}
\hat's large eccentricity and our small uncertainties
lead to a result that deviates by more than $2\sigma$ from Eq.~\ref{Eq:tsp}.

The results derived from Eqs.~\ref{Eq:impact-par}, \ref{Eq:trans-dur}, and \ref{Eq:tsp}
are fully consistent with the orbit of the planet.
During primary transit the planet is closer to the star and moves faster,
leading to a smaller impact parameter and shorter transit duration than for the secondary eclipse.
$T_{14,\mathrm{sec}}$ is almost $24$~minutes longer than the primary transit
with only a small error of roughly $20$~seconds.
The mid-time of the secondary eclipse occurs $3.22$~days after the primary eclipse,
leading to a separation in time almost twice as long as that between secondary and primary ($1.67$~days);
the uncertainties for $\Delta t_\mathrm{s-p}$ lie between two and four minutes.

\subsubsection{Light travel time effect}

When star and planet move around their common center of mass,
the planet is physically more distant from the observer during secondary eclipse than during primary eclipse.
Thus, that part of the light coming from the planet just before and after the secondary eclipse must have traveled longer to reach the observer.
Consequently, the secondary transit is observed later than it actually occurs geometrically.
In particular, it is delayed by
\begin{equation}
\Delta t_\mathrm{travel} \, = \, \frac{2 a}{c} \frac{m_s^2 - m_p^2}{\left( m_s + m_p \right)^2} \frac{1-e^2}{1 - e^2 \sin^2 \omega} \; ,
\label{Eq:speed-of-light}
\end{equation}
where  $c$ is the speed of light,
and \mbox{$m_s = 0.809^{+0.020}_{-0.027} \,M_\odot$} and \mbox{$m_p = (0.081 \pm 0.009)\,M_J$} are the masses of star and planet taken from \cite{Bakos2010}.

Using Eq.~\ref{Eq:speed-of-light}, we estimate the amplitude of the light travel time effect to be \mbox{$\Delta t_\mathrm{travel} \approx 44$~s}
for our orbital solution. Given a temporal binning of seven minutes for the secondary eclipse data and an uncertainty of two to four minutes in our mid-transit time for the secondary,
we estimate that considering $\Delta t_\mathrm{travel}$ in our results would change the eccentricity and the argument of periastron by only about a third of their uncertainty interval, slightly moving $e$ to smaller and $\omega$ to larger values.  Thus
light travel time effects play only a minor role in our modeling.

\subsection{Albedo and equilibrium temperature}

The measured depth of the secondary eclipse
\begin{equation}
d_s \, = \, \frac{1}{1+f_3} \left( \frac{R_p}{R_s} \right)^2
\label{Eq:sec-trans-depth}
\end{equation}
provides an opportunity to determine the albedo and the equilibrium temperature of \hat.
The secondary eclipse depth is related to the planetary geometric albedo~$A_g$ through
\begin{equation}
d_s \, = \, A_g \cdot \left( \frac{R_p}{r_\mathrm{sec}} \right)^2 \; ,
\label{Eq:geo-albedo}
\end{equation}
where we use the distance $r_\mathrm{sec}$ of the planet from the star during secondary eclipse
instead of the semi-major axis~$a$ to account for the eccentric orbit.

A potentially confounding factor in relating the secondary eclipse depth to the geometric albedo is 
a possible contribution of thermal emission to the eclipse depth.
The fraction of the planetary thermal flux~$f_\mathrm{thermal}$ in the \textit{Kepler} bandpass, which
covers wavelengths between $\lambda_1 \approx 4000$~\AA\ and $\lambda_2 \approx 9000$~\AA,
can be estimated by
\begin{equation}
f_\mathrm{thermal} \, = \, \frac{\pi \int^{\lambda_2}_{\lambda_1} B_\lambda (T_\mathrm{eq}) \, d\lambda}{\sigma_\mathrm{SB} T_\mathrm{eq}^4} \, \left( \frac{a}{R_s} \right)^2
\label{Eq:f-thermal}
\end{equation}
from \citet{Han2014}, where $T_\mathrm{eq}$ is the planetary equilibrium temperature,
$B_\lambda$ is the blackbody intensity, and $\sigma_\mathrm{SB}$ is the Stefan-Boltzmann constant. 
In the following calculations we show that this contribution is negligible.

The planetary equilibrium temperature
is related to the stellar effective temperature~$T_s$, the Bond albedo~$A_b$,
and the redistribution factor~$f_\mathrm{redist}$, characterizing the energy redistribution within
the planetary atmosphere, as
\begin{equation}
T_\mathrm{eq}(t) \, = \, T_s \cdot \left( f_\mathrm{redist} \frac{R_s}{r(t)} \right)^\frac{1}{2} \, \left( 1 \, - \, A_b \right)^\frac{1}{4} \ ,
\label{Eq:eq-temp}
\end{equation}
where the factor $R_s/r(t)$ accounts for the distance of the planet from the bright stellar surface.
An energy redistribution factor~$f_\mathrm{redist}$ of $2/3$ accounts for a non-uniform distribution of
flux over the surface of the planet with the substellar point receiving the most intense irradiation;
$f_\mathrm{redist}=1/2$ represent the case when all incoming flux
is reradiated isotropically from each point of the irradiated hemisphere \citep[see][]{Hansen2008}.
A uniform distribution of flux over the entire surface of the planet would be accounted for by $f_\mathrm{redist}=1/4$.

Following \citet{Han2014}, we assume that geometric and Bond albedo are related via
$A_b = 3/2 \, A_g$.
Due to the non-vanishing eccentricity, the planetary equilibrium temperature becomes a function of
the time-dependent orbital distance~$r(t)$ of the planet, which we use in Eq.~\ref{Eq:eq-temp}.
In this expression, any dynamical effects in the planetary atmosphere are of course ignored and instantaneous
adjustment of the atmosphere to the received radiative input is assumed.

\begin{figure}[t]
  \includegraphics[width=0.45\textwidth]{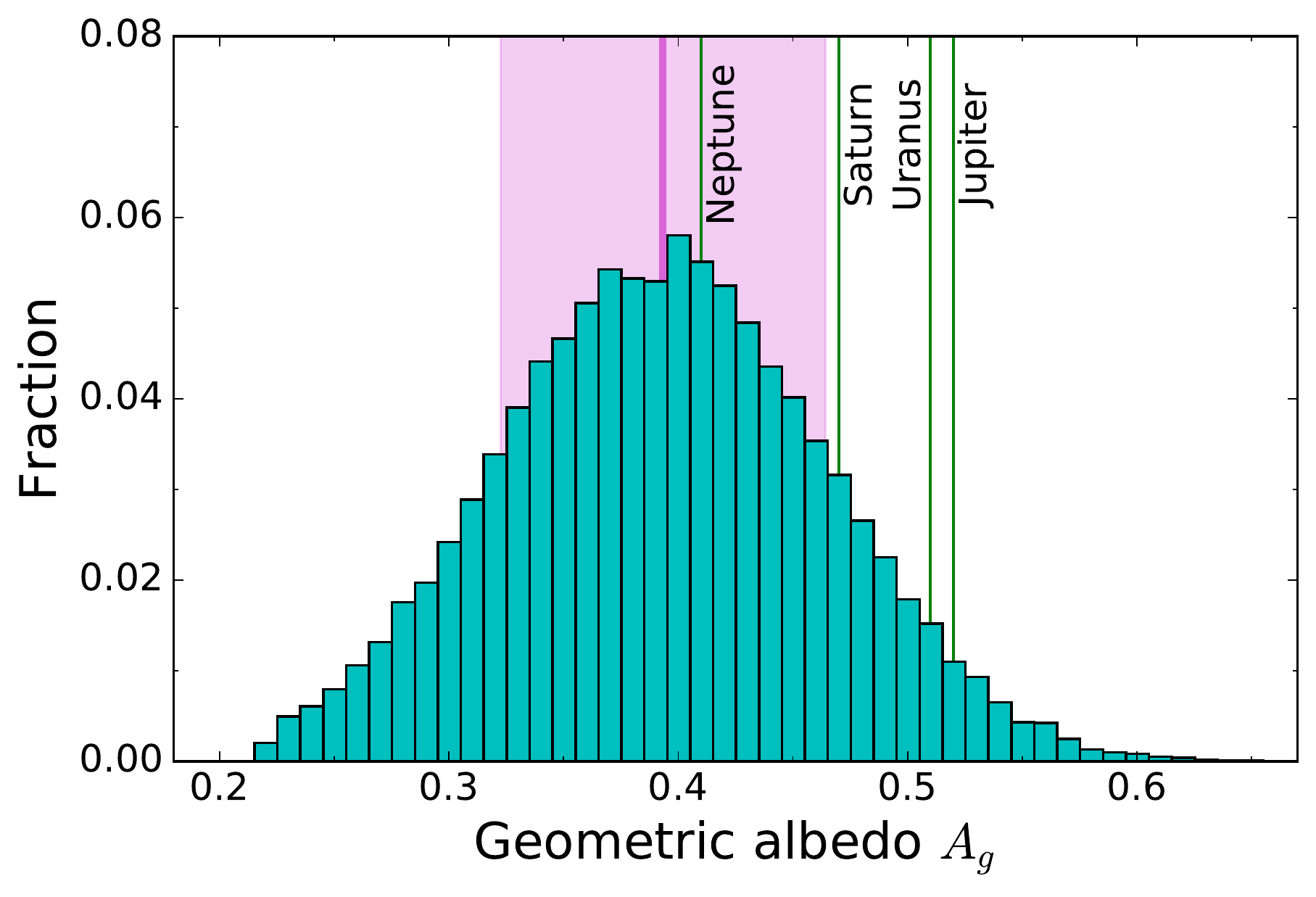}
  \caption{Distribution of the geometric albedo $A_g$ of \hat\ (histogram).
           The (magenta) line and area indicate the median and $68$~\% credibility interval.
           The visual geometric albedos of the Solar System giant planets are given for comparison.
  \label{fig:albedo}}
\end{figure}

Adopting $T_s = (4780\pm50)$~K \citep{Bakos2010} as the effective stellar temperature of HAT-P-11,
we calculate the geometric albedo~$A_g$ and planetary effective temperature~$T_\mathrm{eq}$ based on
the traces of our MCMC sampling with priors.
In Fig.~\ref{fig:albedo} we present the marginal distribution for the geometric albedo~$A_g$,
including the visual geometric albedos of the Solar System giant planets\footnote{\url{http://nssdc.gsfc.nasa.gov/planetary/factsheet/}}
for reference.

We ignore the thermal correction to $A_g$ because
$f_\mathrm{thermal}$ is only $< 2\cdot10^{-4}$ for $f_\mathrm{redist} = 2/3$, and becomes even smaller for $f_\mathrm{redist} = 1/2$, which is negligible.
Even if we insert the highest~$T_\mathrm{eq}$ of the planet on its orbit, $f_\mathrm{thermal}$ is still insignificant.
Although we would have to measure the phase curve of \hat\ to derive reliable estimates of the brightness of the planet's night side,
the low estimate for the fraction of thermal flux indicates that the
night side of \hat \ remains virtually invisible to \textit{Kepler}, which records
only the reflected stellar flux.

Figure~\ref{fig:albedo} shows that Neptune's albedo lies well inside the credibility interval of \mbox{$A_g = 0.39\pm0.07$},
close to the expectation value.
The albedo of Saturn is only slightly beyond the $84$~\% quantile and those of Uranus and Jupiter are about $2\sigma$ off.
The planet \hat\ has roughly the size of Neptune and also a similar albedo, which might indicate a further similarity between these two bodies, even though their orbits are very different.

In recent years several albedos of exoplanets have been measured using \textit{Kepler} data. 
\citet{Angerhausen2015} present a comprehensive study of $20$~confirmed \textit{Kepler} planets.
They detect secondary eclipses in $16$ cases and derive geometric albedos which are all smaller than our result for \hat,
although the albedo of Kepler-44~b (\mbox{KOI-204}) has a relatively large uncertainty.
Their largest value is \mbox{$A_g = 0.32\pm0.03$} for Kepler-7~b (KOI-97), a low-density planet with
a mass of $M_p = 0.43 \, M_J$ and a radius of $R_p = 1.48 \, R_J$ \citep{Latham2010}.
\citet{Esteves2015} study a sample of $14$~\textit{Kepler} planets reporting that $11$ objects have albedos smaller than $0.25$.
They argue that from the three cases with larger albedos probably only the result of Kepler-10~b with \mbox{$A_g = 0.58\pm0.25$} is reliable;
interestingly, Kepler-10~b \citep{Bat2011} is a rather special planet in their sample with a very short period, small radius,
and rocky composition, probably not having an atmosphere.
Furthermore, \citet{Sheets2014} published an analysis of $31$~sub-Saturn \textit{Kepler} candidate planets
and determine the average albedo of the sample.
Excluding \mbox{Kepler-10~b} from their sample, they find a geometric albedo of \mbox{$A_g = 0.22\pm0.06$},
whereas \mbox{Kepler-10~b} represents an outlier with \mbox{$A_g = 0.60\pm0.09$}.
It seems that \mbox{\hat} lies at the upper end of geometric albedos measured for exoplanets with the radius of Neptune or larger;
it maybe even occupies a niche between these planets and super-Earths such as \mbox{Kepler-10~b} with rocky compositions and thin or no atmospheres.
However, Kepler-10~b might also just represent a statistical outlier; after all the geometric albedos of Solar System rocky planets are small,
except for Venus with its dense atmosphere.
Finally, we note that in contrast to \mbox{\hat} none of the exoplanets in the samples of \citet{Sheets2014}, \citet{Esteves2015}, and \citet{Angerhausen2015}
shows a significant eccentricity.
Whether this has an effect on the albedo, however, remains to be determined.

\begin{figure}[t]
  \includegraphics[width=0.45\textwidth]{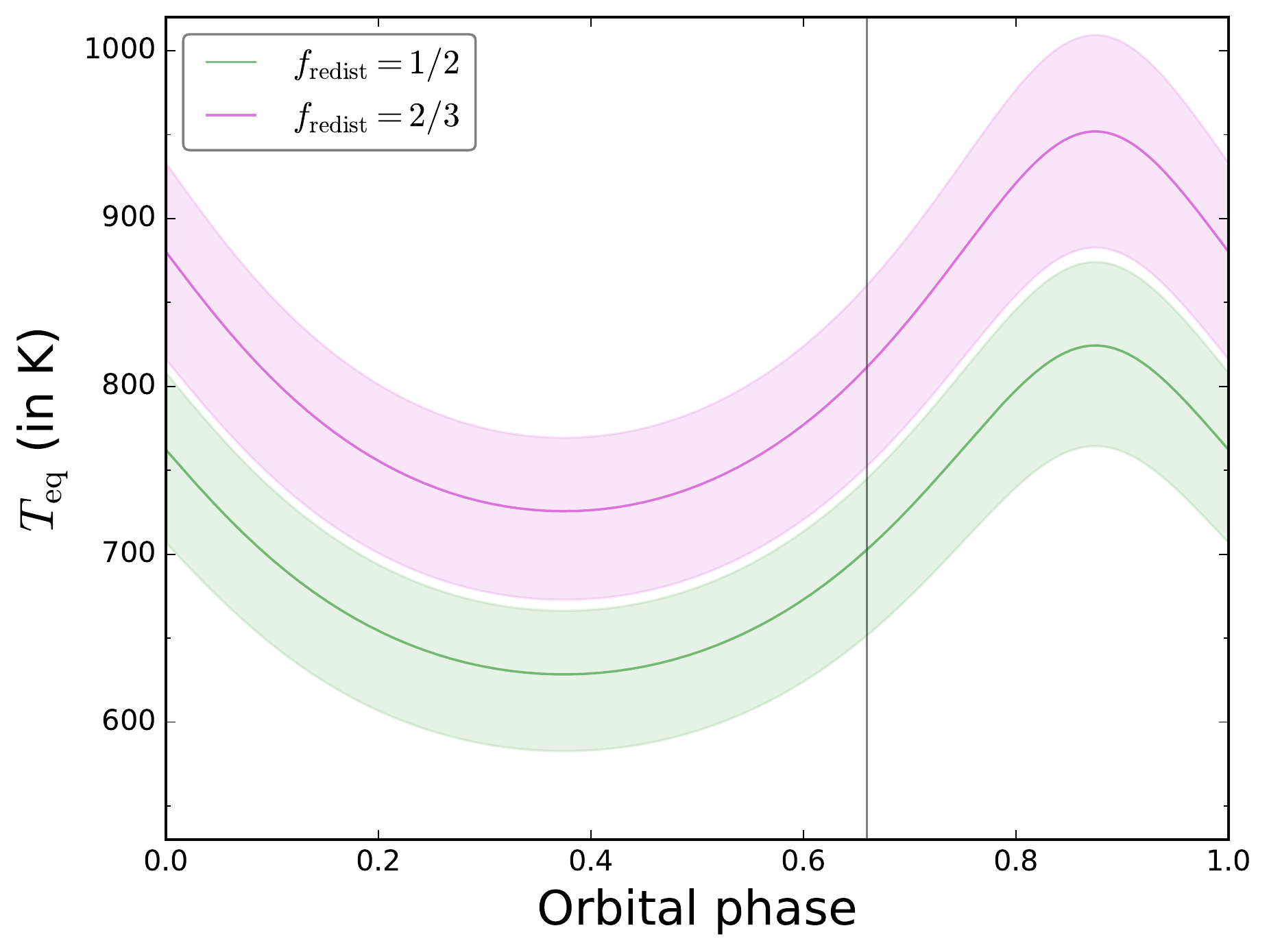}
  \caption{Equilibrium temperature $T_\mathrm{eq}$ of \hat\ over orbital phase for
           a redistribution factor $f_\mathrm{redist}$ of $1/2$ and $2/3$.
           The vertical line is the position of the secondary eclipse.
           The shaded areas indicate the $68$~\% credibility intervals.
  \label{fig:Teq}}
\end{figure}

We show the derived equilibrium temperature of \hat\ in Fig.~\ref{fig:Teq};
the colored margins represent a full error propagation for all input parameters to Eq.~\ref{Eq:eq-temp}, with the exception of $r(t)$, which has a negligible contribution.
The dominating uncertainty comes from the stellar effective temperature~$T_s$.
We emphasize that $T_\mathrm{eq}$ represents an estimate for the brightness temperature
of \hat's day side, which is seen close to secondary eclipse.
The temperature on the night side depends crucially on the energy transport 
mechanisms in the planet's
atmosphere and the rotation of the planet, both of which remain unknown.
Thus, Fig.~\ref{fig:Teq} should not be confused with the ``brightness'' of the visible hemisphere of \hat\ over its orbital phase.

$T_\mathrm{eq}$ is a function of time and highest when the planet is closest to the star near its periastron.  Because at that time the planet also moves fastest, 
this ``hot phase'' of the planet lasts only for a relatively
small period of time compared to the entire orbital period.
Thus, on the far side of the orbit the equilibrium temperatures are lowest and the temperature gradient is smallest as well.
Since we do not know the redistribution of heat in the planetary atmosphere,
we present the temperature for both $f_\mathrm{redist}=2/3$ and $f_\mathrm{redist}=1/2$.
As expected from Eq.~\ref{Eq:eq-temp}, the values for $f_\mathrm{redist}=2/3$ are higher.
Both models differ the least ($\Delta T \approx 100^\circ$~K) when the planet is coolest,
but the temperature difference rises to almost $130^\circ$~K during periastron.
The median temperature of the planet for $f_\mathrm{redist}=1/2$ is $680^\circ$~K, with a
maximum temperature difference of $200^\circ$~K over one orbit;
for $f_\mathrm{redist}=2/3$ the corresponding values are $790^\circ$~K and $230^\circ$~K.
Even for the latter high-temperature solution, \hat\ remains cooler than the day side of Mercury over half of its orbital period.

\section{Summary}
\label{sec:Summary}

We present an in-depth transit analysis of all available \textit{Kepler} short-cadence data of the
planetary system \mbox{HAT-P-11}.
First, we derived updated planetary ephemeris;
using our revised reference epoch $T_0$ and orbital period $P_p$,
we were able to report the first detection of the secondary eclipse of \hat\ based on the phase-folded \textit{Kepler} light curve.
Second, we carried out simultaneous primary and secondary transit modeling.
Here, we specifically selected a sample of primary transit light curves which are only weakly affected by spot-crossing signatures.
The orbital elements and planetary parameters determined from our modeling are consistent with previous estimates for the primary transit,
although we present significantly reduced statistical uncertainties for most of the parameters.

In combination with the secondary eclipse, we pin down the eccentric orbit of \hat\
to a much higher precision than previously determined using RV measurements;
the phasing of the secondary eclipse yields an orbital eccentricity of $0.26459_{-0.00048}^{+0.00069}$.
We determine a secondary eclipse depth of $6.09_{-1.11}^{+1.12}$~ppm, which translates into a
geometric planetary albedo of $0.39\pm0.07$ for \hat.
Among the Solar System giant planets, this is best consistent with Neptune;
however, the uncertainty remains too large to reliably rule out consistency with the albedos of the other giants.
Whether the similarity between \hat\ and Neptune reaches beyond their sizes and albedos has to be determined by future studies.

Due to \hat's non-circular orbit, the planetary equilibrium temperature changes with phase.
We determine temperatures between $630^\circ$~K and $950^\circ$~K, which depend on assumptions on the atmospheric energy redistribution;
no detailed modeling of the atmosphere has been attempted here.
While our analysis does not provide direct information on the night side brightness of \hat,
we estimate the fraction of planetary thermal flux in the \textit{Kepler} bandpass to be small.
In combination with the strong rotational modulation of the light curve due to starspots,
this makes the study of the phase curve of this planet in this data set highly challenging.
However,
future observations at infrared wavelengths, where the impact of stellar activity is weaker (e.g., by SOFIA or JWST), might allow to resolve the planetary phase curve and provide further insight into the physics of this intriguing planetary system.

\begin{acknowledgements}
All of the \textit{Kepler} data presented in this paper were obtained from the Mikulski Archive for Space Telescopes (MAST).
STScI is operated by the Association of Universities for Research in Astronomy, Inc., under NASA contract
NAS5-26555. Support for MAST for non-HST data is provided by the NASA Office of Space Science via
grant NNX09AF08G and by other grants and contracts.
This paper includes data collected by the Kepler mission. Funding for the Kepler
mission is provided by the NASA Science Mission directorate.
This research has made use of the SIMBAD database, operated at CDS, Strasbourg, France \citep{Wenger2000}.
This research has made use of the Exoplanet Orbit Database
and the Exoplanet Data Explorer at exoplanets.org \citep{Han2014}.
This work made extensive use of PyAstronomy\footnote{\url{https://github.com/sczesla/PyAstronomy}}.
KFH thanks the German Research Foundations (DFG) for grants under HU 2177/1-1.
\end{acknowledgements} 

\bibliographystyle{aa}
\bibliography{doc.bib}

\end{document}